# Wavelength Spacing Tunable, Multiwavelength Q-Switched Mode-Locked Laser Based on Graphene-Oxide-Deposited Tapered Fiber

Lei Gao, Tao Zhu, *Member, IEEE,* and Jing Zeng

*Abstract*—A wavelength spacing tunable, multiwavelength Q-switched mode-locked (QML) fiber laser in an erbium-doped fiber cavity based on graphene oxide deposited on tapered fiber is proposed by choosing the diameter and length of the taper, graphene oxide thickness and cavity dispersion, in which the wavelength spacing could be tuned by pump power. The evolutions of temporal and spectral with different pump strengths are investigated. Results show that the tunability of the multiwavelength laser can be interpreted by the bound states of QML laser resulting from a mutual interaction of dispersion, nonlinear effect, insertion loss, and pump power. To the best of our knowledge, it is the first experimental observation of bound states of QML, which provides a new mechanism to fabricate tunable multiwavelength laser.

## I. INTRODUCTION

TUNABLE multiwavelength mode-locked fiber lasers (MLs) have drawn considerable attentions for their various applications, such as wavelength-division multiplexing, optical sensing and microwave/terahertz generation [1-3]. Generally, mode-locking (ML) operation can be classified as continuous wave mode-locking (CWML) and Q-switched mode-locking (QML) [3]. Laser functioning in CWML shows uniform pulses and their repetition frequency usually matches with the laser cavity round time, while the temporal output under QML displays a giant Q-switched envelope above the pulse train and two characteristic repetition frequencies are show in their frequency spectrum (RF). Compared to CWML, QML would produce pulses with higher energy, which is very importance in micro-fabrication, nonlinear frequency conversion [4]. To date, several techniques have been proposed for QML fiber lasers based on saturable absorber (SA) effect, including nonlinear polarization rotation (NPR) [5], single wall carbon nanotubes [6], semiconductor saturable absorber mirrors [7]. However, they all suffer large loss, bulk size, complex fabrication or narrow wavelength range. In contrast, SAs based on graphene have shown remarkable merits for ML fiber laser [8-9]. As a two-dimensional layer of hexagonal packed carbon atoms, graphene shows excellent SA effect due to the Pauli blocking caused by the linear dispersion of the Dirac electrons, including broad wavelength range, large modulation depth and low saturation intensity [10]. But, pristine graphene has no appreciable solubility in most solvent, making it hard for manipulation. Recently, researches shown that graphene oxide (GO) has outstanding SA effect and also good solubility, making it ideal for ML fiber laser [11].

GO is synthesized by oxidation of graphite, and the functional groups, like hydroxyl and epoxide functional groups on their sheet edges, make it disperse well in water. Together with various structures for integrating GO-based saturable absorber (GOSA) into laser cavity, ML pulses with different shapes, periods and center wavelengths have been reported [11-13]. Nevertheless, due to the homogeneous gain medium of erbium-doped fiber (EDF) at room temperature, stable multiwavelength laser is difficult to achieve under strong mode competition, and additional component has to be added to achieve stable multiwavelength ML pulses, including cooling EDF in liquid nitrogen [2], in-fiber comb filters [14], or specific fiber gratings [15-16], which lead to high cost and bulk size. Based on graphene, He propose a wavelength-tunable ML via adjusting chirped fiber Bragg grating [17], and Luo produce several multiwavelength ML with filtering effect and four wave mixing (FWM) [18-20]. Besides, Ahmad demonstrate a eleven wavelengths Brillouin Q-switch fiber laser and a tunable spectrum ML with a Mach-Zehnder filter [21, 22]. Yet, their wavelength spacing are fixed, and the wavelength tuning is inconvenient.

In this letter, we proposed a wavelength spacing tunable, multiwavelength QML fiber laser in an EDF cavity by exploiting GOSA deposited on tapered fiber. By rotating polarization controller (PC), stable QML is generated, and a stable multiwavelength QML fiber lasers with different spectral shapes can be obtained by tuning pump power. Results show that this multiwavelength laser can be explained by the bound states of QML laser, which requires a delicate balance of dispersion, nonlinear effects and pump power.

## II. EXPERIMENTAL SETUP

The GO used in our experiment is prepared by chemical oxidation, and its microscope image is depicted in Fig. 1 (a). Figure 1 (b) shows its Raman spectroscopy, and the D-band and

This work was supported by Natural Science Foundation of China (No. 61377066), the Fundamental Research Funds for the Central Universities (No. CDJZR12125502 and 106112013CDJZR120002).

Lei Gao, Tao Zhu, Jing Zeng are with the Key Laboratory of Optoelectronic Technology & Systems (Education Ministry of China), Chongqing University, Chongqing 400044, China. (Corresponding email: zhutao@cqu.edu.cn)

G-band appear at around 1319 cm$^{-1}$ and 1606 cm$^{-1}$, respectively. The high ratio of $I_D/I_G$ less than 1.04 indicates that the GO has a large amount of defects. Then, 0.5 mg GO is immersed into 100 mL N,N-dimethylformamide solution, and after 20 minutes of ultra-sonication, the solution is centrifuged to get uniform and transparent layer. In order to get fiber taper with low loss and good repeatability, we use a optical fiber coupler machine with a standard single mode fiber (SMF, Corning SMF-28) [23]. During the tapering process, an optical spectrum analyzer (OSA, Si720, Micro Optics) is utilized to monitor the transmission spectrum of the fiber taper. The waist diameter of the taper used in this paper is ~ 6.5 μm, and the waist length is ~5 mm. The result shown in Fig. 1 (c) indicates an insertion loss of 1.5 dB. After immersing the fiber taper into a droplet of transparent GO solution, a 200 mW continuous wave 980 nm laser is injected into one port of the taper, and the output power is monitored by an optical power meter connecting to the other port. Two minutes later, the GO deposition loss is ~7 dB, then the taper is removed from GO solution and fixed in a clean box for natural drying. Such a high loss guarantees enough GO flakes to function with the evanescent field. The transmission spectrum after deposition shows no absorption peak, indicating the broad absorption range. The image of the taper after deposition is shown in Fig. 1 (d), apparently that the GO flakes with a thickness of ~1.5 μm have been well deposited on the taper waist.

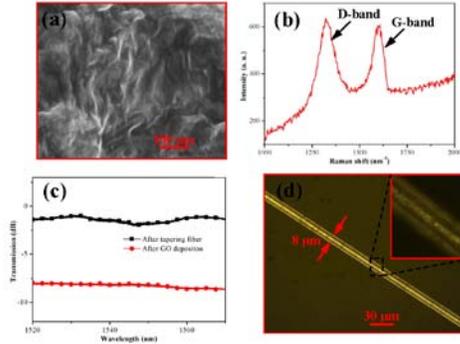

Fig. 1. (a) Microscope image of GO; (b) Raman spectrum of GO samples excited with a 632.8 nm laser; (c) transmission spectra of the fiber taper before and after GO deposition; (d) microscope image of GO deposited on tapered fiber.

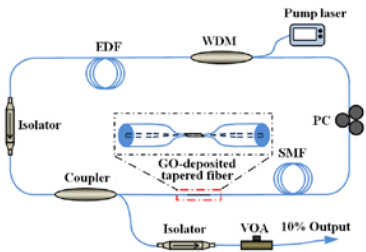

Fig. 2. Schematic diagram of the fiber ring laser.

The scheme of fiber ring cavity incorporating the GOSA fabricated above is shown in Fig. 2. A 10 m long EDF (EDF-980-T2, Stockeryale, Inc.) with a peak absorption of 6.56 dB/m at 1530 nm is used as gain medium, which is pumped by a 980 nm laser through a wavelength division multiplexer (WDM). A polarization-independent isolator is used for unidirectional operation, and a PC is employed to change the birefringence of the cavity. The output laser is extracted from a 10% fiber coupler. Other fibers used in the cavity are standard SMF (SMF-28, Corning Inc.). The total length of the ring cavity is about 24.5 m, corresponding to a fundamental frequency of ~8.4 MHz. The group velocity dispersions (GVD) parameters of EDF and SMF are -12.7 and 18 ps/nm/km, respectively, corresponding to the total anomalous GVD as 0.0155 ps$^2$.

During the experiment, the temporal output of laser is monitored by a detector (PDB430C, 350MHz, Thorlabs Co,. Ltd) and visualized by a real time oscilloscope (Infiniium MSO 9404A, Agilent Tech.) and a frequency analyzer (DSA815, Rigol Tech.). An OSA (B6142B, Agilent Tech.) is utilized to measure its optical spectra. To ensure the detector is not saturated, a variable optical attenuator (VOA) is inserted between the detector and output.

### III. EXPERIMENTAL RESULTS AND DISCUSSION

*A. Characteristics of the QML Laser*

By increasing the pump power from 0 to 380 mW with a rate of ~5 mW, both QML and CWML fiber lasers are observed and represented in Fig. 3, and the temporal output train of QML is shown in Fig. 4 (a). As shown, the temporal output under QML shows a giant Q-switched envelope above the neatly spaced short pulse trains. As expected, the RF spectra of QML in Fig. 4 (b) contain two characteristic repetition frequencies of 25.6 kHz and 8.42 MHz, demonstrating its QML characteristic.

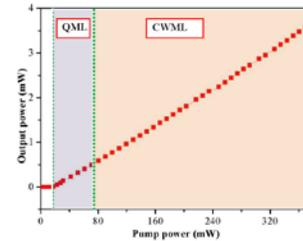

Fig. 3. Total output powers under different pump powers.

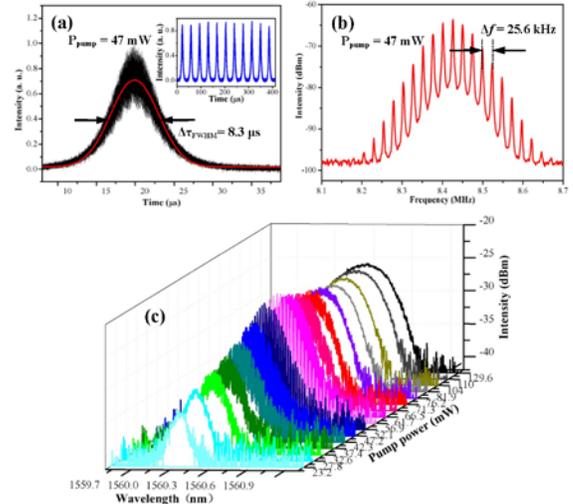

Fig. 4. Characteristics of QML lasers. (a) Temporal pulse train with pump power at 47 mW, fitted with Gauss profile, the inset shows the pulse train in a large range. (b) RF spectra of output. (c) Optical spectra under different pump powers.

*B. Pump-dependent Characteristics*

The spectra under different pump powers are depicted in Fig. 4 (c). As shown, the multiwavelength QML lasing emerges once the pump power exceeds 32.6 mW, and its wavelength spacing increases as pump power below 53 mW, then decreases when further increasing pump power. The wavelength spacing of the multiwavelength vs. pump strength is plotted in Fig. 5 (a), which is very sensitive to pump strength. In our experiment, a maximum wavelength spacing of 0.145 nm is recorded, corresponding to a minimum wavelength number of 7. However, QML lasers with much less wavelength number are rationally expected when pumped with a smaller pumping step. As for pump power larger than 76.2 mW, this multiwavelength QML laser disappears, but a typical CWML laser with single wavelength emerges and lasting for pump power in the range from 76.2 mW to 380 mW.

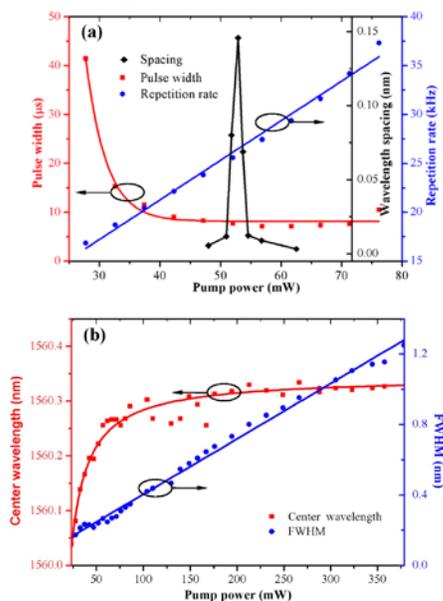

Fig. 5. Pump-dependent characteristics of QML lasers. (a) Temporal pulse width and repetition rate versus pump power. The pulse width is fitted exponentially while repetition rate is fitted linearly; (b) The center wavelength and the FWHM of lasing wavelength versus pump power. The center wavelength is fitted exponentially while FWHM is fitted linearly.

The pulse width and pulse repetition rate are plotted as a function of pump power in Fig. 5 (a). The repetition rate increases linearly from 16.8 kHz to 37.3 kHz while pulse width decreases from 41.5 μs to 7 μs when pump power increases from 20 mW to 76 mW. Further increasing pump power results in CWML fiber laser. The laser center wavelength and the full width at half maximum (FWHM) versus pump strength are shown in Fig. 5 (b), where the envelope is counted for the multiwavelength region. Results show that the center wavelength red-shifts for larger pump power, and the wavelength FWHM increases linearly with the increment of pump power.

*C. Bound States of QML*

Conventionally, the multi-peak modulation of Q-switched pulses is observed in passively Q-switched fiber lasers, where both a single pulse and the pulse trains are modulated by a fundamental cavity frequency [24-26]. This phenomenon is also referred as self-mode-locking, which originates from the intermodal beats. Yet, results in in our experiment is different from the previous observations for pulse modulation period in our scheme can be different greatly in a small pumping region. We attribute the multiwavelength laser to bound state (BS) in QML fiber laser, and five typical optical spectra are shown in Fig. 6, corresponding BSs with different orders.

Different from previous reports [6, 13], no continuous wave laser ever shown before QML, and QML emerges other than CWML as long as the laser is formed. The former is mainly induced by the low saturation energy of GOSA, while the later is mainly caused by the fact that a minimum intra-cavity pulse energy to bleaching the GOSA is required to get CWML fiber laser [27], and the larger loss of GOSA would also expedite the tendency toward QML [4]. Noting that this BSs are found in QML region, so it is BS of QML rather than CWML. Limited by the maximum bandwidth (350 MHz) of our recorded system, their corresponding temporal trains are obscure, yet when the wavelength spacing is relative small, the difference of their temporal outputs can be noticed, and we give the spectra and the corresponding temporal trains for some pump powers in Figs. 7 (a)-(c). The experimental results clearly show that BSs with different pulse separations lead to multiwavelength QML lasers with diverse spectral spacing.

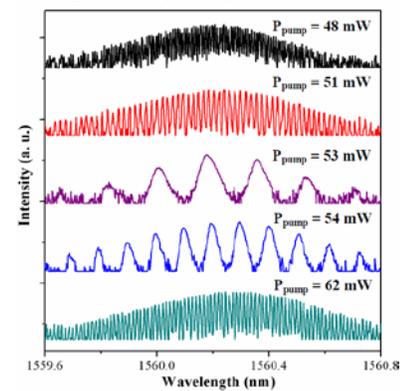

Fig. 6 Optical spectra for pump power at 48 mW, 51 mW, 53 mW, 54 mW and 61.7 mW, respectively.

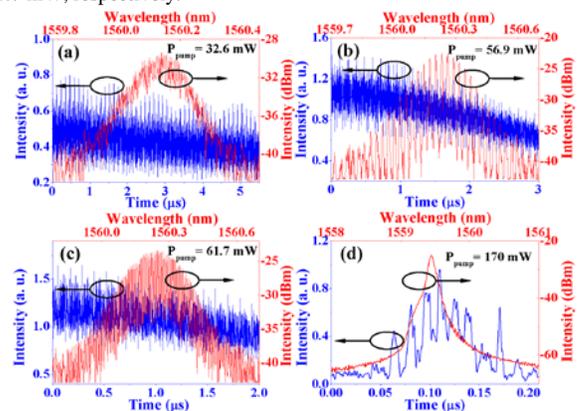

Fig. 7 (a)-(c) BSs of QML for pump power at 32.6 mW, 56.9 mW and 61.7 mW, respectively; (d) BS of CWML in cavity with additional 20 m of DCF.

Previously, a number kinds of BSs have been observed in CWML fiber laser based on NPR, figure-eight-cavity and other real SAs [28-35]. As the solutions described by the complex

Ginzburg-Landau equation, the formation of BS in CWML mainly considered as the results of a balance of repulsive and attractive forces caused by dispersion and nonlinear effects [28-29]. However, due to the complexity of the laser dynamics, the numerical solution of complex Ginzburg-Landau equation describing BS of QML is unknown. Yet, considering the pump-dependence of QML, it is possible to obtain BS of QML when cavity parameters are carefully selected. Here, we find that the BS can also formed in QML laser, which requires a more delicate balance of mutual forces within the cavity.

The formation of BSs in QML requires a delicate balance of dispersion, nonlinear effect, insertion loss and most of importance, pump strength. First, the BS of QML can be only found in cavity with net anomalous GVD, which is verified by splicing different lengths of SMF and dispersion compensating fiber (DCF38, Thorlabs Co,. Ltd) into the cavity to change its net GVD from anomalous to normal. BSs of QML exist only in cavity with net anomalous GVD, and BS of CWML can be found in that with net normal GVD, which can be shown in Fig. 7 (d) with small wavelength spacing and large modulation period. Second, we fabricate GOSAs with various deposition losses, namely GO thickness, and experiments indicate that BS of QML cannot be find in cavities with GOSA deposition loss less than ~5 dB or larger than ~10 dB. Namely, a proper insertion loss of the SA is very important in generating BS of QML. Besides, the waist diameter of the fiber taper has to be optimized. All of the results suggest the total insertion loss is vital important to the formation of BS of QML. Third, considering that QML is formed under low pump strength, it is not surprising to find out that the BSs in QML fiber laser are extremely sensitive to pump power. Different from BSs in CWML that higher pump strength would bring higher order BSs due to soliton energy quantization [27], BSs in QML require an optimal pump power for each types of BSs, and the pump power larger or smaller than the optimum value would change the BS order severely. This intensely pump-dependent characteristic can be easily understood that energy in the laser cavity is insufficient for completely bleaching the GOSA, so a slight variation of pump power would lead to great change of mutual forces between the laser pulses.

### D. Potential Applications

Considering that the wavelength spacing of the QML can be tuned extremely flexible via pump strength in less than 5 mW, our scheme would find potential applications in wavelength division multiplexing (WDM) [36, 37] and FWM [19, 38]. Besides, the frequency spacing and microwave signal period can be controlled with facility in a very short pump region, which is very useful in optical frequency metrology and tunable microwave generation [39-41].

## IV. CONCLUSION

we proposed a tunable multiwavelength QML fiber laser in an EDF cavity based on GOSA, and its temporal, spectral characteristics under different pump powers are discussed. Results show that the wavelength number can be tuned by pump intensity, of which can be explained by the formation of BSs in QML laser that resulting from a mutual interaction of dispersion, nonlinear effects, insertion loss and pump power. To our best knowledge, this letter is the first report about experimental observation of BSs of QML laser, which provides a new mechanism to fabricate tunable multiwavelength laser. This kind of tunable multiwavelength QML laser is beneficial to understanding the complex dynamics of laser physics, and also would find potential applications in optical communication, sensing and microwave/terahertz generation.

**Lei Gao** was born in Henan, China, in 1989. He received the B.E. degree in Opto-electronic Information Engineering from Chongqing University, China, in 2011. He is currently working toward the Ph.D. degree in Optical Engineering from Chongqing University, Chongqing, China. His current research interests include optical fiber passively mode-locked laser, and nonlinear optics.

**Tao Zhu** (M'06) received the Ph.D. degree in Optical Engineering from Chongqing University, China, in 2008. During 2008-2009, he worked in Chongqing University, China. During 2010-2011, he is a Postdoctoral Research Fellow at the Department of Physics in University of Ottawa, Canada. Since April 2011, he is a professor of Chongqing University, China. He has published over 80 papers in the international journals and the conference proceedings. His research focuses on passive and active optical components, optical sensors, and distributed optical fiber sensing system.

Dr. Zhu is a member of the IEEE and the Optical Society of America.

**Jing Zeng** was born in Guizhou, China, in 1988. He received the B.S. degree in Physics from Yangtze University, China, in 2012. He is currently working toward the M.S. degree in Optical Engineering from Chongqing University, Chongqing, China. His current research is optical fiber laser.